\documentclass[aps,prb,showpacs,showkeys,twocolumn,preprintnumbers,
               amsmath,amssymb,floatfix,
					citeautoscript,
					superscriptaddress]{revtex4}
\usepackage{graphicx}	

\begin{document}

\preprint{}

\title{Piezoelectric exciton-acoustic phonon coupling in single quantum dots}

\author{D.~Sarkar}
\email{dipankar.sarkar@uam.es}

\affiliation{Departamento de F\'{i}sica de Materiales, Universidad Aut\'{o}noma de Madrid, E-28049 Madrid, Spain}
\author{H.~P.~van~der~Meulen}
\affiliation{Departamento de F\'{i}sica de Materiales, Universidad Aut\'{o}noma de Madrid, E-28049 Madrid, Spain}
\author{J.~M.~Calleja}
\affiliation{Departamento de F\'{i}sica de Materiales, Universidad Aut\'{o}noma de Madrid, E-28049 Madrid, Spain}
\author{J.~M.~Meyer}
\affiliation{Institut f\"ur Festk\"{o}rperphysik, Leibniz Universit\"{a}t Hannover, D-30167 Hannover, Germany}
\author{R.~J.~Haug}
\affiliation{Institut f\"ur Festk\"{o}rperphysik, Leibniz Universit\"{a}t Hannover, D-30167 Hannover, Germany}
\author{K.~Pierz}
\affiliation{Physikalisch-Technische Bundesanstalt Braunschweig, D-38116 Braunschweig, Germany}

\date{ \today}

\begin{abstract}

Micro-photoluminescence spectroscopy at variable temperature, excitation intensity and energy was performed on a single InAs/AlAs self-assembled quantum dot. The exciton emission line (zero-phonon line, ZPL) exhibits a broad sideband due to exciton-acoustic phonon coupling by the deformation potential mechanism. Additionally, narrow low-energy sidebands at about 0.25 meV of the ZPL are attributed to exciton-acoustic phonon piezoelectric coupling. In  lowering the excitation energy or intensity these bands gradually dominate the emission spectrum of the quantum dot while the ZPL disappears. At high excitation intensity the sidebands due to piezoelectric coupling decrease strongly and the ZPL dominates the spectrum as a consequence of screening of the piezoelectric coupling by the photocreated free carriers.

\end{abstract}   


\maketitle

Semiconductor quantum dots (QDs) are promising candidates for optical devices, both for traditional and for quantum information applications. 
The knowledge of interactions of confined carriers with the environment is of fundamental importance for the device operation because they determine the optical efficiency on one hand and the coherence properties on the other. 
Phase coherence can be lost either by relaxation processes, namely radiative and phonon assisted recombination or by coupling to acoustic phonons (pure dephasing).
The former produces a broadening of the Lorentzian zero-phonon line (ZPL). 
Coupling to acoustic phonons gives rise to non-Lorentzian sidebands. \cite{besombes01}

Temperature induced broadening of the ZPL associated to exciton-phonon scattering was reported for  natural QDs in a GaAs quantum well \cite{fan98} and CdSe/ZnCdSe QDs \cite{gindele99}. 
Acoustic phonon sidebands were observed in CdTe/ZnTe QDs \cite{besombes01,moehl04}, InAs/GaAs QDs \cite{favero03}, GaAs monolayer fluctuation QDs \cite{peter04}, and CdSe QDs\cite{arians07}.

Exciton-acoustic phonon interactions have been treated theoretically considering both the deformation potential (DP) and piezoelectric (PZ) coupling mechanisms \cite{duke65,besombes01,vasilevskiy04prb}. 
It was pointed out theoretically \cite{takagahara93} and experimentally \cite{besombes01} that in small QDs the DP coupling mechanism dominates over PZ coupling. 
However, effects related to piezoelectricity in QDs are still under investigation, as they are very sensible to QD geometry, size and composition.  \cite{stier99,bester05,bester06prl,schliwa07}

In this work we performed micro-photoluminescence (PL) spectroscopy on a single InAs/AlAs QD and measured the influence of temperature, excitation power and excitation energy on the emission spectra.
Besides a broad acoustic phonon sideband, similar to previously reported ones originating from DP coupling \cite{besombes01,favero03,moehl04,peter04,arians07} we observe a narrow low-energy sideband and show that it originates from the PZ exciton coupling to long-wavelength acoustic phonons. 
This interpretation is supported by the strong reduction of the PZ sidebands observed on increasing excitation power, as a consequence of screening by the photocreated carriers.  

The sample was grown by molecular beam epitaxy and self-assembly of InAs dots between 20~nm thick AlAs layers. A 10~nm thick GaAs cap layer was deposited on top of the sample.  
Rotation of the wafer was interrupted during growth of the QD layer, resulting in a density gradient across the wafer and in particular a low density area. The typical sizes of the dots measured by atomic force microscopy before capping are of the order of 20~nm in diameter and 2 to 3~nm in height. 
Additionally aluminum masks were fabricated on top of the sample by e-beam lithography and lift-off with openings of 0.2~to 10~$\mu$m in order to access optically single QDs.

The sample was mounted in a continuous-flow cryostat for microscope applications and cooled with helium down to 10~K. 
A microscope objective with numerical aperture of 0.55 focused the exciting laser beam on a $\mu$m-sized spot and also collected the signal. 
Argon and Ti:Sapphire lasers were used for excitation. 
Detection was done with a LN$_2$ cooled Si-CCD mounted on a 0.85~m double grating spectrometer.

In Fig.~1 we show the PL spectra of the exciton transition of a single QD for temperatures of 10~and 30~K. We have verified that all peaks belong to the same QD by observing the spectral jitter and the polarization- and power dependence. \cite{turck00} The exciton transition consists of a fine-structure split doublet of linearly cross-polarized peaks x and y. The weak peaks at higher energy are most probably charged exciton complexes of the QD, as shown by jitter and resonant excitation experiments \cite{sarkar08}. The large fine-structure splitting (FSS) of 0.28~meV is due to the high AlAs barriers. \cite{sarkar06} The peaks shift to lower energies with increasing temperature, due to thermal expansion. At 10~K we observe broad acoustic phonon wings (light gray area). They are more pronounced at the low energy side \cite{ikeda06} reflecting the higher probability of emitting than absorbing a phonon.  With increasing temperature this asymmetry is gradually removed and the wing and ZPL broaden as reported for II-VI \cite{besombes01,moehl04,arians07} and III-V \cite{favero03,peter04}  semiconductor QDs.  This wing corresponds to the widely reported \textit{DP mediated} exciton-acoustic phonon interaction. Note, that the FSS of the two exciton transitions is sufficiently large to resolve two separate maxima in the low energy wing at about 1.2~meV (DP arrows) corresponding to the most efficiently coupled acoustic phonon modes. 
This value is larger than the 0.7~meV estimated \footnote{ The strongest coupling is found in Ref. ~\onlinecite{besombes01} for a wave vector  equal to the inverse of the dot size. For a QD size of 9 nm, this gives 0.66~meV. (Fig.~3 of Ref.~\onlinecite{besombes01})}  for CdTe QDs. 
The ratio between these values is 1.7. Assuming similar QD sizes, the value for this ratio can be estimated from the ratio of the corresponding sound velocities $v_{\text{LA}}$. 
For QD materials the ratio is 1.3 
($v_{\text{LA,InAs}} = 4.28~\cdotp 10^5$~cm/s,
 $v_{\text{LA,CdTe}} = 3.34~\cdotp 10^5$~cm/s)\cite{adachi05book},
while for barrier materials the ratio is 1.6 
($v_{\text{LA,AlAs}} = 6.24~\cdotp 10^5$~cm/s,
 $v_{\text{LA,ZnTe}} = 3.84~\cdotp 10^5$~cm/s)\cite{adachi05book}.


\begin{figure}[t!]  
  \includegraphics[width=8cm]{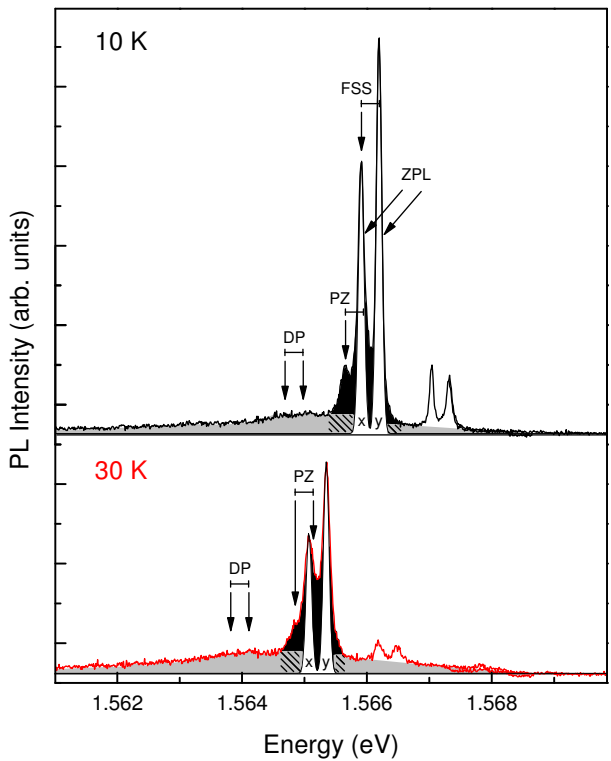}
  \caption{(Color online) Photoluminescence spectra of a fine-structure split (FSS) exciton doublet (x, y) of a single quantum dot for temperatures of 10 and 30~K, exciting at 2.41~eV. The doublet is composed of a central Lorentzian zero phonon line (ZPL) and asymmetric acoustic phonon wings.} 
  \label{fig1}
\end{figure}  

 \begin{figure}[t!]  
  \includegraphics[width=8cm]{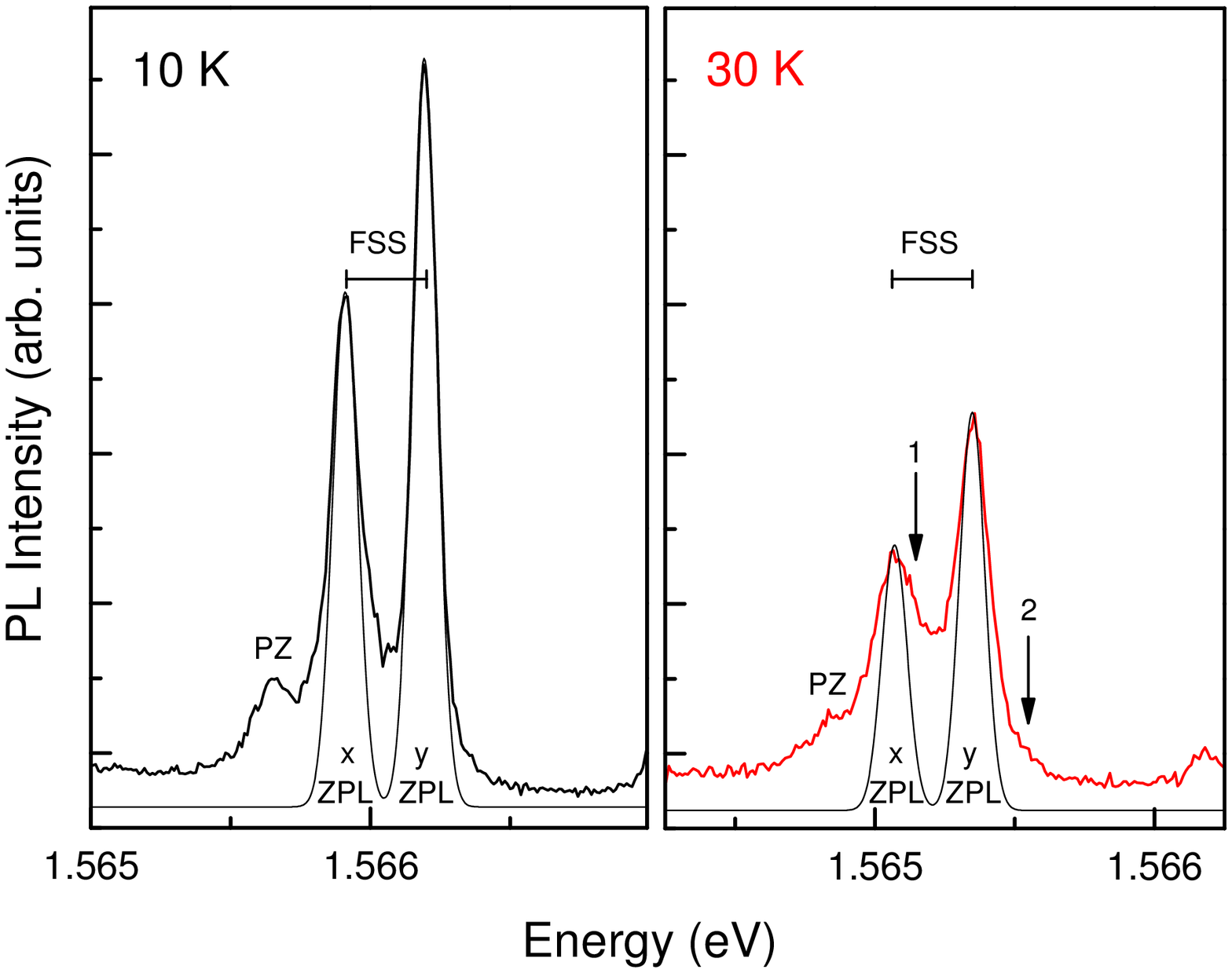}
  \caption{(Color online) The narrow sidebands due to  piezoelectric (PZ) exciton-acoustic phonon coupling are shown. Left panel (10~K): The PZ sideband of the x-peak is observed  0.2~meV below the zero phonon line (ZPL). The sideband corresponding to the y-ZPL results in an additional broadening of the x-peak. Right panel (30~K): By increasing the temperature sidebands also appear on the high energy sides of the y-ZPL (arrow~2) and x-ZPL (arrow~1, perceived as elevated x-y valley).} 
  \label{fig2}
\end{figure}  

Additional features (black areas) are observed close to the ZPL, which are the main focus of this work. They are shown in detail in Fig.~2. At 10~K we observe an additional narrow low-energy band located at 0.25~meV below the x-ZPL (marker PZ). A similar band can be perceived below the y-ZPL: it is masked by the x-ZPL, giving rise to an apparent additional broadening of the x peak. By increasing the temperature (30~K), a wing appears on the high energy side of the y-ZPL (arrow~2).  The corresponding wing for the x-ZPL (arrow~1) results in an elevation of the x-y valley. Thus, concerning temperature dependence the low-energy sidebands behave similar to the DP wings. We attribute this narrow low-energy sidebands to the interaction of the exciton with \textit{long wavelength} acoustic phonons mediated by \textit{PZ coupling}.
 
In order to support this assignment, we performed excitation intensity dependent measurements, shown in Fig.~3. The spectra are normalized to the excitation power, which ranges from $ P_0 /160 $ to $ P_0 $ ($P_0 = 0.4$~mW). The left and right panel show PL spectra for excitation below (1.98~eV) and above (2.14~eV) the wetting layer absorption edge, respectively. For decreasing excitation power the acoustic phonon wings become wider and more pronounced while the ZPLs lose weight compared to the sidebands. For the lowest intensities the ZPLs are not even observed any longer and the emission lineshape is completely dominated by the PZ sidebands. This power dependence of the sidebands and ZPLs is indicative of screening. In contrast to the DP coupling mechanism, PZ coupling can be screened out by free carriers \cite{duke65,takagahara93}. By increasing the excitation power, more electron-hole plasma is generated and  the PZ mediated exciton-acoustic phonon interaction is screened. When exciting above the wetting layer absorption edge (right panel in Fig.~3) the screening is more efficient than exciting below (left panel). Therefore the power dependence observed in Fig.~3 confirms the attribution to the PZ coupling mechanism.

The PZ sidebands are much closer ($E_{PZ} = 0.25$~meV) to the ZPL than the corresponding DP wings ($E_{DP} = 1.2$~meV). One possible explanation of that difference is the wave vector dependence of both coupling mechanisms:
The electron-acoustic phonon coupling mediated by DP interaction increases with the wave vector $q$ as $\sqrt{q}$ while the unscreened PZ mediated coupling goes as \cite{duke65} $1/\sqrt{q}$. This additional $1/q$ dependence of the PZ coupling results in an stronger interaction for small $q$ (long-wavelength) acoustic phonons \cite{yu01book}.

\begin{figure}[t!]  
  \includegraphics[width=8cm]{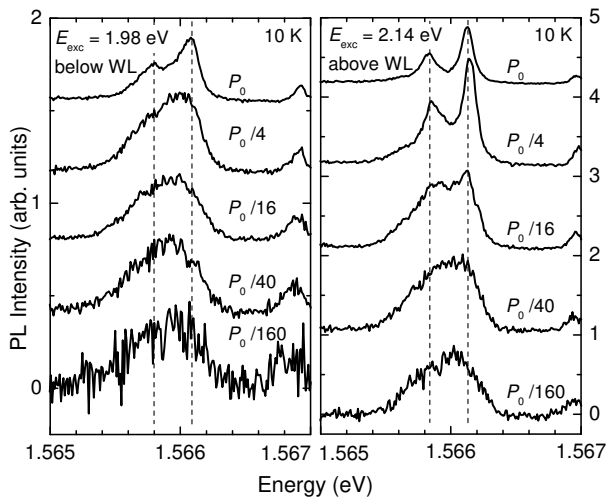}
  \caption{Photoluminescence spectra of the exciton doublet for a series of excitation power. Left and right panel show the spectra for excitation below and above the wetting layer (WL). For increasing excitation intensities the sidebands get less pronounced opposite to the zero phonon line, which gains weight. This indicates a screening effect of the acoustic phonon-exciton coupling.  Spectra are taken at 10~K.} 
  \label{fig3}
\end{figure}  

From the energy maxima of the sidebands we can estimate the QD size and screening length.
The energy of the DP sideband maxima are related to the QD size by $E_{DP} = \hbar v_{\text{LA}} / \xi$, $ \xi$ being the localization length \cite{takagahara85,besombes01}. This gives a rough estimate of around 10~nm for our QD, consistent with the atomic force microscopy measurements. On the other hand, the maximum of the PZ sideband allows us to estimate the Debye screening length $\lambda_D=1 /k_D  \approx \hbar v_s / E_{PZ} \approx 10$ nm ($v_s = 2550$~m/s is the mixed LA/TA sound velocity \cite{duke65}), which is of the order of the QD size. This confirms that screening is relevant in our system. Indeed we can estimate the density of mobile photocreated charges: $n_0 = \epsilon \epsilon_0 k_B T / e^2 \lambda_D^2 \approx 10^{16}$~cm$^{-3}$, with temperature $T=10$~K and static dielectric constant $\epsilon = 15$. $k_B$ is the Boltzmann constant, $\epsilon_0$ is the vacuum dielectric constant and $e$ the elementary charge.	

Figure~4 exhibits the total apparent linewidth of the exciton peak versus the excitation energy for constant excitation intensity. We observe a line narrowing for increasing excitation energy. It appears more pronounced for the x exciton because it overlaps with the wing of the y peak. This is in contrast to experiments with InGaAs QDs \cite{kammerer02prb} where line narrowing is observed for decreasing excitation energy and is interpreted in terms of electrostatic broadening: A reduced excess energy $ E_{excit} - E_X $ leads to less spectral diffusion due to fewer charges in the vicinity of the QD. Thus, we exclude jitter as the origin of the exciton broadening in our case.  Line narrowing with increasing excitation energy was reported for InAlGaAs QDs\cite{leosson03}, and tentatively attributed to increasingly efficient evaporation of excess carriers from the QD, but the origin of the effect is left open. In our case, we attribute the line narrowing to a more efficient creation of free carriers at energies above the wetting layer, which shield the exciton PZ coupling to acoustic phonons. Therefore we conclude that charges photocreated in the wetting layer are responsible for the screening.

\begin{figure}[t!]  
  \includegraphics[width=8cm]{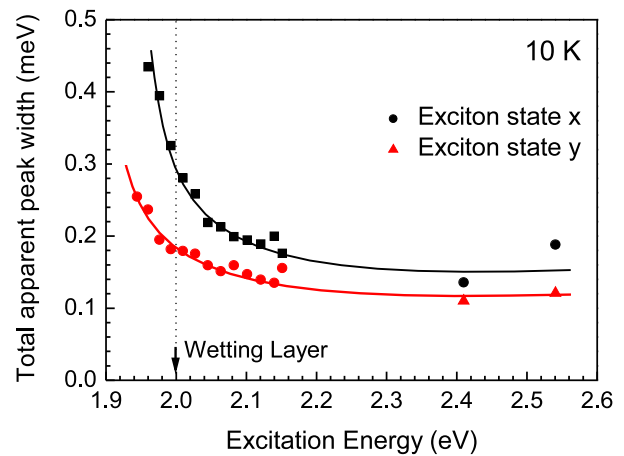}
  \caption{(Color online) Total apparent linewidth of the exciton peaks versus excitation energy for constant excitation power. A transition to broader peaks can be observed for excitation  below the wetting layer. Continuous lines are drawn as guide to the eye.} 
  \label{fig4}
\end{figure}  

The large FSS \cite{sarkar06} and the power/energy dependent exciton-acoustic phonon coupling manifest strong PZ effects in our dots compared to previously reported experiments. In general QDs present different strain distributions, which depend delicately on the QD geometry and give rise to PZ effects. Both linear \cite{stier99,bester05} and particularly quadratic piezoelectric effects  \cite{bester06prl,schliwa07} are very sensitive to QD shape, size and composition, so that both contributions might even cancel out \cite{bester06prb}. Regarding the composition, part of the effect can be attributed to the larger PZ coefficient $ e_{14} =$ -0.23$~C/m^2 $ (bulk, estimated) of AlAs compared to GaAs (-0.16) and InAs (-0.045) (bulk, experimental) \cite{adachi05book}. For GaAs and InAs the calculated effect of strain increases \cite{bester06prl} $ e_{14} $ to  -0.230 (GaAs) and -0.115 (InAs). For strained AlAs also an even higher value might be expected.  Additionally, intermixing \cite{offermans05, sarkar08} of aluminum will rise the PZ coefficient of the InAs QD. Referring to the QD geometry, it has been predicted that a high vertical aspect ratio \cite{schliwa07}, which allows a large vertical separation of the electron and hole charges, increases the PZ effects. The QD under investigation are the larger ones, emitting at the low energy tail of the QD ensemble. It was reported that for InAs/GaAs \cite{saito99} and InAs/AlAs QDs \cite{ballet99} a large QD volume involves also a large vertical aspect ratio and consequently would lead to stronger PZ effect. Thus, both morphology and composition of our QDs definitely can lead to stronger PZ effects compared to InAs/GaAs ones.

In summary, we show that piezoelectric effects play a important role in InAs/AlAs QDs. Particularly, a considerable contribution to the exciton-long wavelength acoustic phonon interaction is mediated by PZ coupling. The interaction can be screened be photo-created extrinsic carriers, and this switching ability may be used in future quantum devices.

This work has been supported by research contracts of the Spanish Ministry of Education (MEC MAT2005-01388, NAN2004-09109-C04-04, TEC2004-05260-C02-02, Consolider CSD 2006-19) and the Community of Madrid (CAM S-0505-ESP-0200).

\bibliography{paper}

\begin{thebibliography}{27}
\expandafter\ifx\csname natexlab\endcsname\relax\def\natexlab#1{#1}\fi
\expandafter\ifx\csname bibnamefont\endcsname\relax
  \def\bibnamefont#1{#1}\fi
\expandafter\ifx\csname bibfnamefont\endcsname\relax
  \def\bibfnamefont#1{#1}\fi
\expandafter\ifx\csname citenamefont\endcsname\relax
  \def\citenamefont#1{#1}\fi
\expandafter\ifx\csname url\endcsname\relax
  \def\url#1{\texttt{#1}}\fi
\expandafter\ifx\csname urlprefix\endcsname\relax\def\urlprefix{URL }\fi
\providecommand{\bibinfo}[2]{#2}
\providecommand{\eprint}[2][]{\url{#2}}

\bibitem[{\citenamefont{Besombes et~al.}(2001)\citenamefont{Besombes, Kheng,
  Marsal, and Mariette}}]{besombes01}
\bibinfo{author}{\bibfnamefont{L.}~\bibnamefont{Besombes}},
  \bibinfo{author}{\bibfnamefont{K.}~\bibnamefont{Kheng}},
  \bibinfo{author}{\bibfnamefont{L.}~\bibnamefont{Marsal}}, \bibnamefont{and}
  \bibinfo{author}{\bibfnamefont{H.}~\bibnamefont{Mariette}},
  \bibinfo{journal}{Phys. Rev. B} \textbf{\bibinfo{volume}{63}},
  \bibinfo{pages}{155307} (\bibinfo{year}{2001}).

\bibitem[{\citenamefont{Fan et~al.}(1998)\citenamefont{Fan, Takagahara,
  Cunningham, and Wang}}]{fan98}
\bibinfo{author}{\bibfnamefont{X.}~\bibnamefont{Fan}},
  \bibinfo{author}{\bibfnamefont{T.}~\bibnamefont{Takagahara}},
  \bibinfo{author}{\bibfnamefont{J.~E.} \bibnamefont{Cunningham}},
  \bibnamefont{and} \bibinfo{author}{\bibfnamefont{H.}~\bibnamefont{Wang}},
  \bibinfo{journal}{Solid State Commun.} \textbf{\bibinfo{volume}{108}},
  \bibinfo{pages}{857} (\bibinfo{year}{1998}).

\bibitem[{\citenamefont{Gindele et~al.}(1999)\citenamefont{Gindele, Hild,
  Langbein, and Woggon}}]{gindele99}
\bibinfo{author}{\bibfnamefont{F.}~\bibnamefont{Gindele}},
  \bibinfo{author}{\bibfnamefont{K.}~\bibnamefont{Hild}},
  \bibinfo{author}{\bibfnamefont{W.}~\bibnamefont{Langbein}}, \bibnamefont{and}
  \bibinfo{author}{\bibfnamefont{U.}~\bibnamefont{Woggon}},
  \bibinfo{journal}{Phys. Rev. B} \textbf{\bibinfo{volume}{60}},
  \bibinfo{pages}{R2157} (\bibinfo{year}{1999}).

\bibitem[{\citenamefont{Moehl et~al.}(2004)\citenamefont{Moehl, Tinjod, Kheng,
  and Mariette}}]{moehl04}
\bibinfo{author}{\bibfnamefont{S.}~\bibnamefont{Moehl}},
  \bibinfo{author}{\bibfnamefont{F.}~\bibnamefont{Tinjod}},
  \bibinfo{author}{\bibfnamefont{K.}~\bibnamefont{Kheng}}, \bibnamefont{and}
  \bibinfo{author}{\bibfnamefont{H.}~\bibnamefont{Mariette}},
  \bibinfo{journal}{Phys. Rev. B} \textbf{\bibinfo{volume}{69}},
  \bibinfo{pages}{245318} (\bibinfo{year}{2004}).

\bibitem[{\citenamefont{Favero et~al.}(2003)\citenamefont{Favero, Cassabois,
  Ferreira, Darson, Voisin, Tignon, Delalande, Bastard, Roussignol, and
  Gerard}}]{favero03}
\bibinfo{author}{\bibfnamefont{I.}~\bibnamefont{Favero}},
  \bibinfo{author}{\bibfnamefont{G.}~\bibnamefont{Cassabois}},
  \bibinfo{author}{\bibfnamefont{R.}~\bibnamefont{Ferreira}},
  \bibinfo{author}{\bibfnamefont{D.}~\bibnamefont{Darson}},
  \bibinfo{author}{\bibfnamefont{C.}~\bibnamefont{Voisin}},
  \bibinfo{author}{\bibfnamefont{J.}~\bibnamefont{Tignon}},
  \bibinfo{author}{\bibfnamefont{C.}~\bibnamefont{Delalande}},
  \bibinfo{author}{\bibfnamefont{G.}~\bibnamefont{Bastard}},
  \bibinfo{author}{\bibfnamefont{P.}~\bibnamefont{Roussignol}},
  \bibnamefont{and} \bibinfo{author}{\bibfnamefont{J.~M.}
  \bibnamefont{Gerard}}, \bibinfo{journal}{Phys. Rev. B}
  \textbf{\bibinfo{volume}{68}}, \bibinfo{pages}{233301}
  (\bibinfo{year}{2003}).

\bibitem[{\citenamefont{Peter et~al.}(2004)\citenamefont{Peter, Hours,
  Senellart, Vasanelli, Cavanna, Bloch, and Gerard}}]{peter04}
\bibinfo{author}{\bibfnamefont{E.}~\bibnamefont{Peter}},
  \bibinfo{author}{\bibfnamefont{J.}~\bibnamefont{Hours}},
  \bibinfo{author}{\bibfnamefont{P.}~\bibnamefont{Senellart}},
  \bibinfo{author}{\bibfnamefont{A.}~\bibnamefont{Vasanelli}},
  \bibinfo{author}{\bibfnamefont{A.}~\bibnamefont{Cavanna}},
  \bibinfo{author}{\bibfnamefont{J.}~\bibnamefont{Bloch}}, \bibnamefont{and}
  \bibinfo{author}{\bibfnamefont{J.~M.} \bibnamefont{Gerard}},
  \bibinfo{journal}{Phys. Rev. B} \textbf{\bibinfo{volume}{69}},
  \bibinfo{pages}{041307} (\bibinfo{year}{2004}).

\bibitem[{\citenamefont{Arians et~al.}(2007)\citenamefont{Arians, Kummell,
  Bacher, Gust, Kruse, and Hommel}}]{arians07}
\bibinfo{author}{\bibfnamefont{R.}~\bibnamefont{Arians}},
  \bibinfo{author}{\bibfnamefont{T.}~\bibnamefont{Kummell}},
  \bibinfo{author}{\bibfnamefont{G.}~\bibnamefont{Bacher}},
  \bibinfo{author}{\bibfnamefont{A.}~\bibnamefont{Gust}},
  \bibinfo{author}{\bibfnamefont{C.}~\bibnamefont{Kruse}}, \bibnamefont{and}
  \bibinfo{author}{\bibfnamefont{D.}~\bibnamefont{Hommel}},
  \bibinfo{journal}{Appl. Phys. Lett.} \textbf{\bibinfo{volume}{90}},
  \bibinfo{pages}{101114} (\bibinfo{year}{2007}).

\bibitem[{\citenamefont{Duke and Mahan}(1965)}]{duke65}
\bibinfo{author}{\bibfnamefont{C.~B.} \bibnamefont{Duke}} \bibnamefont{and}
  \bibinfo{author}{\bibfnamefont{G.~D.} \bibnamefont{Mahan}},
  \bibinfo{journal}{Phys. Rev.} \textbf{\bibinfo{volume}{139}},
  \bibinfo{pages}{A1965} (\bibinfo{year}{1965}).

\bibitem[{\citenamefont{Vasilevskiy et~al.}(2004)\citenamefont{Vasilevskiy,
  Anda, and Makler}}]{vasilevskiy04prb}
\bibinfo{author}{\bibfnamefont{M.~I.} \bibnamefont{Vasilevskiy}},
  \bibinfo{author}{\bibfnamefont{E.~V.} \bibnamefont{Anda}}, \bibnamefont{and}
  \bibinfo{author}{\bibfnamefont{S.~S.} \bibnamefont{Makler}},
  \bibinfo{journal}{Physical Review B} \textbf{\bibinfo{volume}{70}},
  \bibinfo{pages}{035318} (\bibinfo{year}{2004}).

\bibitem[{\citenamefont{Takagahara}(1993)}]{takagahara93}
\bibinfo{author}{\bibfnamefont{T.}~\bibnamefont{Takagahara}},
  \bibinfo{journal}{Phys. Rev. Lett.} \textbf{\bibinfo{volume}{71}},
  \bibinfo{pages}{3577} (\bibinfo{year}{1993}).

\bibitem[{\citenamefont{Stier et~al.}(1999)\citenamefont{Stier, Grundmann, and
  Bimberg}}]{stier99}
\bibinfo{author}{\bibfnamefont{O.}~\bibnamefont{Stier}},
  \bibinfo{author}{\bibfnamefont{M.}~\bibnamefont{Grundmann}},
  \bibnamefont{and} \bibinfo{author}{\bibfnamefont{D.}~\bibnamefont{Bimberg}},
  \bibinfo{journal}{Phys. Rev. B} \textbf{\bibinfo{volume}{59}},
  \bibinfo{pages}{5688} (\bibinfo{year}{1999}).

\bibitem[{\citenamefont{Bester and Zunger}(2005)}]{bester05}
\bibinfo{author}{\bibfnamefont{G.}~\bibnamefont{Bester}} \bibnamefont{and}
  \bibinfo{author}{\bibfnamefont{A.}~\bibnamefont{Zunger}},
  \bibinfo{journal}{Phys. Rev. B} \textbf{\bibinfo{volume}{71}},
  \bibinfo{pages}{045318} (\bibinfo{year}{2005}).

\bibitem[{\citenamefont{Bester et~al.}(2006{\natexlab{a}})\citenamefont{Bester,
  Wu, Vanderbilt, and Zunger}}]{bester06prl}
\bibinfo{author}{\bibfnamefont{G.}~\bibnamefont{Bester}},
  \bibinfo{author}{\bibfnamefont{X.}~\bibnamefont{Wu}},
  \bibinfo{author}{\bibfnamefont{D.}~\bibnamefont{Vanderbilt}},
  \bibnamefont{and} \bibinfo{author}{\bibfnamefont{A.}~\bibnamefont{Zunger}},
  \bibinfo{journal}{Phys. Rev. Lett.} \textbf{\bibinfo{volume}{96}},
  \bibinfo{pages}{187602} (\bibinfo{year}{2006}{\natexlab{a}}).

\bibitem[{\citenamefont{Schliwa et~al.}(2007)\citenamefont{Schliwa,
  Winkelnkemper, and Bimberg}}]{schliwa07}
\bibinfo{author}{\bibfnamefont{A.}~\bibnamefont{Schliwa}},
  \bibinfo{author}{\bibfnamefont{M.}~\bibnamefont{Winkelnkemper}},
  \bibnamefont{and} \bibinfo{author}{\bibfnamefont{D.}~\bibnamefont{Bimberg}},
  \bibinfo{journal}{Phys. Rev. B} \textbf{\bibinfo{volume}{76}},
  \bibinfo{pages}{205324} (\bibinfo{year}{2007}).

\bibitem[{\citenamefont{T{\"u}rck et~al.}(2000)\citenamefont{T{\"u}rck, Rodt,
  Stier, Heitz, Engelhardt, Pohl, Bimberg, and Steingr{\"u}ber}}]{turck00}
\bibinfo{author}{\bibfnamefont{V.}~\bibnamefont{T{\"u}rck}},
  \bibinfo{author}{\bibfnamefont{S.}~\bibnamefont{Rodt}},
  \bibinfo{author}{\bibfnamefont{O.}~\bibnamefont{Stier}},
  \bibinfo{author}{\bibfnamefont{R.}~\bibnamefont{Heitz}},
  \bibinfo{author}{\bibfnamefont{R.}~\bibnamefont{Engelhardt}},
  \bibinfo{author}{\bibfnamefont{U.~W.} \bibnamefont{Pohl}},
  \bibinfo{author}{\bibfnamefont{D.}~\bibnamefont{Bimberg}}, \bibnamefont{and}
  \bibinfo{author}{\bibfnamefont{R.}~\bibnamefont{Steingr{\"u}ber}},
  \bibinfo{journal}{Phys. Rev. B} \textbf{\bibinfo{volume}{61}},
  \bibinfo{pages}{9944} (\bibinfo{year}{2000}).

\bibitem[{\citenamefont{Sarkar et~al.}(2008)\citenamefont{Sarkar, van~der
  Meulen, Calleja, Meyer, Haug, and Pierz}}]{sarkar08}
\bibinfo{author}{\bibfnamefont{D.}~\bibnamefont{Sarkar}},
  \bibinfo{author}{\bibfnamefont{H.~P.} \bibnamefont{van~der Meulen}},
  \bibinfo{author}{\bibfnamefont{J.~M.} \bibnamefont{Calleja}},
  \bibinfo{author}{\bibfnamefont{J.~M.} \bibnamefont{Meyer}},
  \bibinfo{author}{\bibfnamefont{R.~J.} \bibnamefont{Haug}}, \bibnamefont{and}
  \bibinfo{author}{\bibfnamefont{K.}~\bibnamefont{Pierz}},
  \bibinfo{journal}{Appl. Phys. Lett.} \textbf{\bibinfo{volume}{92}},
  \bibinfo{pages}{181909} (\bibinfo{year}{2008}).

\bibitem[{\citenamefont{Sarkar et~al.}(2006)\citenamefont{Sarkar, van~der
  Meulen, Calleja, Becker, Haug, and Pierz}}]{sarkar06}
\bibinfo{author}{\bibfnamefont{D.}~\bibnamefont{Sarkar}},
  \bibinfo{author}{\bibfnamefont{H.~P.} \bibnamefont{van~der Meulen}},
  \bibinfo{author}{\bibfnamefont{J.~M.} \bibnamefont{Calleja}},
  \bibinfo{author}{\bibfnamefont{J.~M.} \bibnamefont{Becker}},
  \bibinfo{author}{\bibfnamefont{R.~J.} \bibnamefont{Haug}}, \bibnamefont{and}
  \bibinfo{author}{\bibfnamefont{K.}~\bibnamefont{Pierz}}, \bibinfo{journal}{J.
  Appl. Phys.} \textbf{\bibinfo{volume}{100}}, \bibinfo{pages}{023109}
  (\bibinfo{year}{2006}).

\bibitem[{\citenamefont{Ikeda et~al.}(2006)\citenamefont{Ikeda, Ogawa, Minami,
  Kuroda, and Takita}}]{ikeda06}
\bibinfo{author}{\bibfnamefont{K.}~\bibnamefont{Ikeda}},
  \bibinfo{author}{\bibfnamefont{Y.}~\bibnamefont{Ogawa}},
  \bibinfo{author}{\bibfnamefont{F.}~\bibnamefont{Minami}},
  \bibinfo{author}{\bibfnamefont{S.}~\bibnamefont{Kuroda}}, \bibnamefont{and}
  \bibinfo{author}{\bibfnamefont{K.}~\bibnamefont{Takita}},
  \bibinfo{journal}{Phys. Status Solidi C} \textbf{\bibinfo{volume}{3}},
  \bibinfo{pages}{874} (\bibinfo{year}{2006}).

\bibitem[{\citenamefont{Adachi}(2005)}]{adachi05book}
\bibinfo{author}{\bibfnamefont{S.}~\bibnamefont{Adachi}},
  \emph{\bibinfo{title}{Properties of Group-IV, III-V and II-VI
  Semiconductors}} (\bibinfo{publisher}{John Wiley and Sons},
  \bibinfo{year}{2005}), ISBN \bibinfo{isbn}{0470090324}.

\bibitem[{\citenamefont{Yu and Cardona}(2001)}]{yu01book}
\bibinfo{author}{\bibfnamefont{P.~Y.} \bibnamefont{Yu}} \bibnamefont{and}
  \bibinfo{author}{\bibfnamefont{M.}~\bibnamefont{Cardona}},
  \emph{\bibinfo{title}{Fundamentals of Semiconductors: Physics and Materials
  Properties}} (\bibinfo{publisher}{Springer}, \bibinfo{year}{2001}), ISBN
  \bibinfo{isbn}{3540254706}.

\bibitem[{\citenamefont{Takagahara}(1985)}]{takagahara85}
\bibinfo{author}{\bibfnamefont{T.}~\bibnamefont{Takagahara}},
  \bibinfo{journal}{Physical Review B} \textbf{\bibinfo{volume}{31}},
  \bibinfo{pages}{6552} (\bibinfo{year}{1985}).

\bibitem[{\citenamefont{Kammerer et~al.}(2002)\citenamefont{Kammerer, Voisin,
  Cassabois, Delalande, Roussignol, Klopf, Reithmaier, Forchel, and
  Gerard}}]{kammerer02prb}
\bibinfo{author}{\bibfnamefont{C.}~\bibnamefont{Kammerer}},
  \bibinfo{author}{\bibfnamefont{C.}~\bibnamefont{Voisin}},
  \bibinfo{author}{\bibfnamefont{G.}~\bibnamefont{Cassabois}},
  \bibinfo{author}{\bibfnamefont{C.}~\bibnamefont{Delalande}},
  \bibinfo{author}{\bibfnamefont{P.}~\bibnamefont{Roussignol}},
  \bibinfo{author}{\bibfnamefont{F.}~\bibnamefont{Klopf}},
  \bibinfo{author}{\bibfnamefont{J.~P.} \bibnamefont{Reithmaier}},
  \bibinfo{author}{\bibfnamefont{A.}~\bibnamefont{Forchel}}, \bibnamefont{and}
  \bibinfo{author}{\bibfnamefont{J.~M.} \bibnamefont{Gerard}},
  \bibinfo{journal}{Phys. Rev. B} \textbf{\bibinfo{volume}{66}},
  \bibinfo{pages}{041306} (\bibinfo{year}{2002}).

\bibitem[{\citenamefont{Leosson et~al.}(2003)\citenamefont{Leosson, Birkedal,
  Magnusdottir, Langbein, and Hvam}}]{leosson03}
\bibinfo{author}{\bibfnamefont{K.}~\bibnamefont{Leosson}},
  \bibinfo{author}{\bibfnamefont{D.}~\bibnamefont{Birkedal}},
  \bibinfo{author}{\bibfnamefont{I.}~\bibnamefont{Magnusdottir}},
  \bibinfo{author}{\bibfnamefont{W.}~\bibnamefont{Langbein}}, \bibnamefont{and}
  \bibinfo{author}{\bibfnamefont{J.}~\bibnamefont{Hvam}},
  \bibinfo{journal}{Physica E} \textbf{\bibinfo{volume}{17}},
  \bibinfo{pages}{1} (\bibinfo{year}{2003}).

\bibitem[{\citenamefont{Bester et~al.}(2006{\natexlab{b}})\citenamefont{Bester,
  Zunger, Wu, and Vanderbilt}}]{bester06prb}
\bibinfo{author}{\bibfnamefont{G.}~\bibnamefont{Bester}},
  \bibinfo{author}{\bibfnamefont{A.}~\bibnamefont{Zunger}},
  \bibinfo{author}{\bibfnamefont{X.}~\bibnamefont{Wu}}, \bibnamefont{and}
  \bibinfo{author}{\bibfnamefont{D.}~\bibnamefont{Vanderbilt}},
  \bibinfo{journal}{Phys. Rev. B} \textbf{\bibinfo{volume}{74}},
  \bibinfo{pages}{081305} (\bibinfo{year}{2006}{\natexlab{b}}).

\bibitem[{\citenamefont{Offermans et~al.}(2005)\citenamefont{Offermans,
  Koenraad, Wolter, Pierz, Roy, and Maksym}}]{offermans05}
\bibinfo{author}{\bibfnamefont{P.}~\bibnamefont{Offermans}},
  \bibinfo{author}{\bibfnamefont{P.~M.} \bibnamefont{Koenraad}},
  \bibinfo{author}{\bibfnamefont{J.~H.} \bibnamefont{Wolter}},
  \bibinfo{author}{\bibfnamefont{K.}~\bibnamefont{Pierz}},
  \bibinfo{author}{\bibfnamefont{M.}~\bibnamefont{Roy}}, \bibnamefont{and}
  \bibinfo{author}{\bibfnamefont{P.~A.} \bibnamefont{Maksym}},
  \bibinfo{journal}{Phys. Rev. B} \textbf{\bibinfo{volume}{72}},
  \bibinfo{pages}{165332} (\bibinfo{year}{2005}).

\bibitem[{\citenamefont{Saito et~al.}(1999)\citenamefont{Saito, Nishi, and
  Sugou}}]{saito99}
\bibinfo{author}{\bibfnamefont{H.}~\bibnamefont{Saito}},
  \bibinfo{author}{\bibfnamefont{K.}~\bibnamefont{Nishi}}, \bibnamefont{and}
  \bibinfo{author}{\bibfnamefont{S.}~\bibnamefont{Sugou}},
  \bibinfo{journal}{Appl. Phys. Lett.} \textbf{\bibinfo{volume}{74}},
  \bibinfo{pages}{1224} (\bibinfo{year}{1999}).

\bibitem[{\citenamefont{Ballet et~al.}(1999)\citenamefont{Ballet, Smathers, and
  Salamo}}]{ballet99}
\bibinfo{author}{\bibfnamefont{P.}~\bibnamefont{Ballet}},
  \bibinfo{author}{\bibfnamefont{J.~B.} \bibnamefont{Smathers}},
  \bibnamefont{and} \bibinfo{author}{\bibfnamefont{G.~J.}
  \bibnamefont{Salamo}}, \bibinfo{journal}{Appl. Phys. Lett.}
  \textbf{\bibinfo{volume}{75}}, \bibinfo{pages}{337} (\bibinfo{year}{1999}).

\end{thebibliography}

\end{document}